\documentclass[preprint2]{aastex}
\usepackage{url}
\begin{document}

\title{\large{\rm{ON THE CRUCIAL CLUSTER ANDREWS-LINDSAY 1 \\ AND A 4\% DISTANCE SOLUTION FOR ITS PN}}}

\author{\sc \small D. Majaess$^{1,2}$, G. Carraro$^3$, C. Moni Bidin$^4$, C. Bonatto$^5$, D. Turner$^1$, M. Moyano$^4$, L. Berdnikov$^{6,7}$, E. Giorgi$^{8}$}

\affil{$^1${\footnotesize Department of Astronomy \& Physics, Saint Mary's University, Halifax, NS B3H 3C3, Canada.}}
\affil{$^2${\footnotesize Mount Saint Vincent University, Halifax, NS B3M 2J6, Canada.}}
\affil{$^3${\footnotesize European Southern Observatory, Avda Alonso de Cordova, 3107, Casilla 19001, Santiago, Chile.}}
\affil{$^4${\footnotesize Instituto de Astronom\'ia, Universidad Cat\'olica del Norte, Av.~Angamos 0610, Antofagasta, Chile.}}
\affil{$^5${\footnotesize Departamento de Astronomia, Universidade Federal do Rio Grande do Sul, Av.~Bento Gonalves 9500 Porto Alegre
91501-970, RS, Brazil.}}
\affil{$^6${\footnotesize Moscow M V Lomonosov State University, Sternberg Astronomical Institute, Moscow 119992, Russia.}}
\affil{$^7${\footnotesize Isaac Newton Institute of Chile, Moscow Branch, Universitetskij Pr. 13, Moscow 119992, Russia.}}
\affil{$^{13}${\footnotesize Facultad de Ciencias Astron\'omicas y Geof\'sicas (UNLP), Instituto de Astrof\'isica de La Plata (CONICET, UNLP), Paseo del Bosque s/n, La Plata, Argentina.}}
\email{dmajaess@ap.smu.ca}

\begin{abstract}
Andrews-Lindsay 1 is a pertinent open cluster granted it may host the planetary nebula PHR 1315-6555, yet ambiguities linger concerning its fundamental parameters ($>50$\% scatter).  New multiband $BVJHW_{1-4}$ photometry for cluster and field stars, in concert with observations of recently discovered classical Cepheids, were used to constrain the reddening and velocity-distance profiles along the sight-line. That analysis yielded the following parameters for the cluster: $E(J-H)=0.24\pm0.03$, $d=10.0\pm0.4$ kpc ($d_{JH}=9.9\pm0.6$ kpc, $d_{BV}=10.1\pm0.5$ kpc), and $\log{\tau}=8.90\pm0.15$.  The steep velocity-distance gradient along $\ell \sim 305 \degr$ indicates that two remote objects sharing spatial and kinematic parameters (i.e., PHR 1315-6555 and Andrews-Lindsay 1) are associated, thus confirming claims that the PN is a cluster member (e.g., Parker et al.).  The new distance for PHR 1315-6555 is among the most precise yet established for a Galactic PN ($\sigma/d=4$\%).
\end{abstract}
\keywords{open clusters and associations: individual: Andrews-Lindsay 1 (ESO 96-SC04)---ISM: planetary nebulae: individual: PHR 1315-6555 (PNG 305.3-03.1)}

\section{{\rm \footnotesize INTRODUCTION}}
The motivation behind the search for planetary nebulae (PNe) in star clusters is to secure precise parameters for the former.  Those parameters include the PN's distance, physical dimensions, age, chemical composition, and potentially progenitor mass \citep{pa11,tu11,mb13}.  A solid PN distance could likewise be used to calibrate relationships aimed at establishing distances to field PNe \citep{fr08}.  Reliable distances are particularly desirable since precise parallaxes exist for a mere fraction of the nearest PNe \citep[][$\sigma/\pi\sim5$\%]{be09}, whereas distance estimates for the bulk of PNe display $20-30$\% uncertainties or larger \citep[][the latter's Fig.~6]{st08,gi11}.
Admittedly, the quest to identify numerous cluster PNe has been hampered by several factors.  Massive PNe associated with younger clusters feature short lifetimes ($10^3-10^4$ years).  The matter is compounded by the paucity of old clusters that spawn lower-mass PNe exhibiting longer lifetimes.  Star clusters rarely survive beyond $10^7$ years \citep{bb11}, and the majority which surpass that threshold host evolved constituents that terminate as SNe \citep[e.g.,][their Fig.~2]{ma07}.  Furthermore, the bulk of cataloged Galactic PNe are members of the bulge \citep[][their Fig.~1]{ma07}, which underscores the pertinence of surveys aimed at discovering PNe throughout the Galactic disk where younger clusters reside \citep[][the Macquarie/AAO/Strasbourg H$\alpha$ PN Catalogue, see also \citealt{ja10,lu12}]{pa06}.  

\begin{table*}[!t]
\begin{center}
\small
\caption{Parameters for Andrews-Lindsay 1 (AL1) \label{table1}}
\begin{tabular}{lcllllc}
\hline \hline
Reference & Target & $d$ (kpc) & $E(B-V)$ & $\tau$ (Gyr) & RV (km/s) & [Fe/H]  \\
\hline
\citet{jp94} & AL1 & 7.57 & 0.72 & ... & ... & ... \\
\citet{ca95} & AL1 & 11.8 & 0.75 & 0.7 & ...  & subsolar \\
\citet{cm04} & AL1 & $12\pm1$ & $0.7\pm0.2$ & 0.8 & ...  & ... \\
\citet{fr04} & AL1 & ...  & ... & ... &  $40\pm10$  & $-0.51\pm0.30$ \\
\citet{fr04b} & AL1 & 9.35 & ... & 0.67 & ... & ... \\
\citet{ca05} & AL1 & $16.95^b$ & $0.34\pm0.05$ & $0.8\pm0.2$ & ...  & ... \\
\citet{fr08} & PN & $9.7\pm3.1$ & $0.71^a$ & ... & $51.6\pm15.0$  & ... \\
 & AL1 & ... & ... &  ... & $50\pm10$  & ... \\
\citet{pa11} & AL1 & ... & ... &...  & $57\pm5^a$  & ... \\
 & PN & $10.4\pm3.4$ & $0.83\pm0.08$ &...  & $58\pm2.5$  & ... \\
 & & & & & $59\pm2$  & \\
\hline
\end{tabular} \\  
(a) See text. (b) Observations were undertaken during unsatisfactory weather conditions. 
\end{center}
\end{table*}

Consequently, few promising cases of PNe in Galactic clusters have been reported, and thus any \textit{bona fide} pairs\footnote{Perhaps an  equally important endeavour is to eliminate unreliable calibrators and reputed associations \citep{ma07,ki08,mb13}.} are crucial for various research topics (e.g., constraining the impact of mass loss).  The search for extragalactic cluster PNe has yielded similar results.  \citet{ja13} observed 467 star clusters in M31 to detect PNe that share the former's velocity.  That evaluation enables chance superpositions to be identified, in addition to contamination from unrelated emission sources along the sight-line (e.g., H II regions).   \citet{ja13} concluded that 5 (of 270) globular clusters may host PNe, whereas those targets identified near open clusters likely constitute chance alignments.

A subsample of the tentative Galactic PNe/open cluster pairs includes Abell 8 and Bica 6 \citep{bo08,tu11}, He 2-86 and NGC 4463 \citep{ma07,mb13}, and PHR 1315-6555 and Andrews-Lindsay 1 \citep{pa06,pa11,ma07,fr08}.\footnote{PHR 1315-6555 is likewise designated as PNG 305.3-03.1, and Andrews-Lindsay 1 is cataloged as ESO 96-SC04 and VdB 144.}  Yet an important consideration arises when inferring the progenitor mass of those PNe from single stars near the cluster turnoff.  The canonical hypothesis advocating that PNe stem from single stars does not readily explain their non-spherical morphologies or low formation rate \citep[][see also \citealt{dm13}]{ja13}. Indeed, the detection of PNe within globular clusters, which exhibit turnoff masses below the 1$M_{\sun}$ threshold predicted by PNe models, implies that the progenitor may have been augmented by mass transfer.  Specifically, \citet{ja13} noted that 4 (of 130) Galactic globular clusters host PNe, with the cases split between PNe featuring non-spherical nebulae and high-mass central stars conducive to younger clusters. That evidence, in concert with the realization that three PNe are located in globular clusters hosting numerous X-ray sources, supports claims that multiplicity affects the formation of globular cluster PNe. 

In each of the aforementioned Galactic cases further independent research is required, with a validation standard on par with the magnitude of the discovery.  For example, velocities measured for the PN NGC 2438 by \citet{pk96} indicated it was a cluster member (M46), whereas those by \citet{od63} suggested otherwise.  Independent observations urged by \citet{ma07} and others were subsequently published by \citet{ki08}, and suggested that the pair constitute a chance alignment along the sight-line. Radial velocities for Abell 8 and Bica 6 likewise require confirmation, especially given their potential implications for Galactic dynamics \citep{tu13}. NGC 4463 may host a rare young PN (He 2-86) that exhibits sizable internal extinction \citep[][their comprehensive discussion in \S 5]{mb13}, and the cluster is pertinent for stellar evolution research.  Specifically, NGC 4463 may host a F-supergiant (post onset of core helium burning), the PN, and a massive blue straggler \citep{al07}.

Continuing the cluster PNe project spearheaded by \citet{mb13}, this study aims to bolster the link tying PHR 1315-6555 to Andrews-Lindsay 1, namely by resolving the faint cluster's presently ambiguous fundamental parameters (Table~\ref{table1}).  New $BVJHW_{1-4}$ observations were analyzed to achieve that objective.  $W_{1-4}$ are the four WISE mid-infrared passbands \citep{wr10}.

\section{{\rm \footnotesize ANALYSIS}}
\subsection{{\rm \footnotesize PRIOR STUDIES CONCERNING AL1}}
Distance estimates cited for Andrews-Lindsay 1 span a factor of two \citep[$7.57-16.95$ kpc,][]{jp94,ca05}.  A similar spread exists for the extinction ($E_{B-V}=0.35-0.75$).  The ambiguity arises from the challenging nature of assessing a distant, faint, and reddened cluster that is superposed upon a field which introduces significant contamination \citep[][their Fig.~1]{ca95}.  However, a consensus exists concerning the cluster age ($\tau \sim 0.8$ Gyr), which was partly inferred from the relative positions of the cluster's red clump and blue turnoff stars.

\begin{figure}[!t]
\begin{center}
\includegraphics[width=7.5cm]{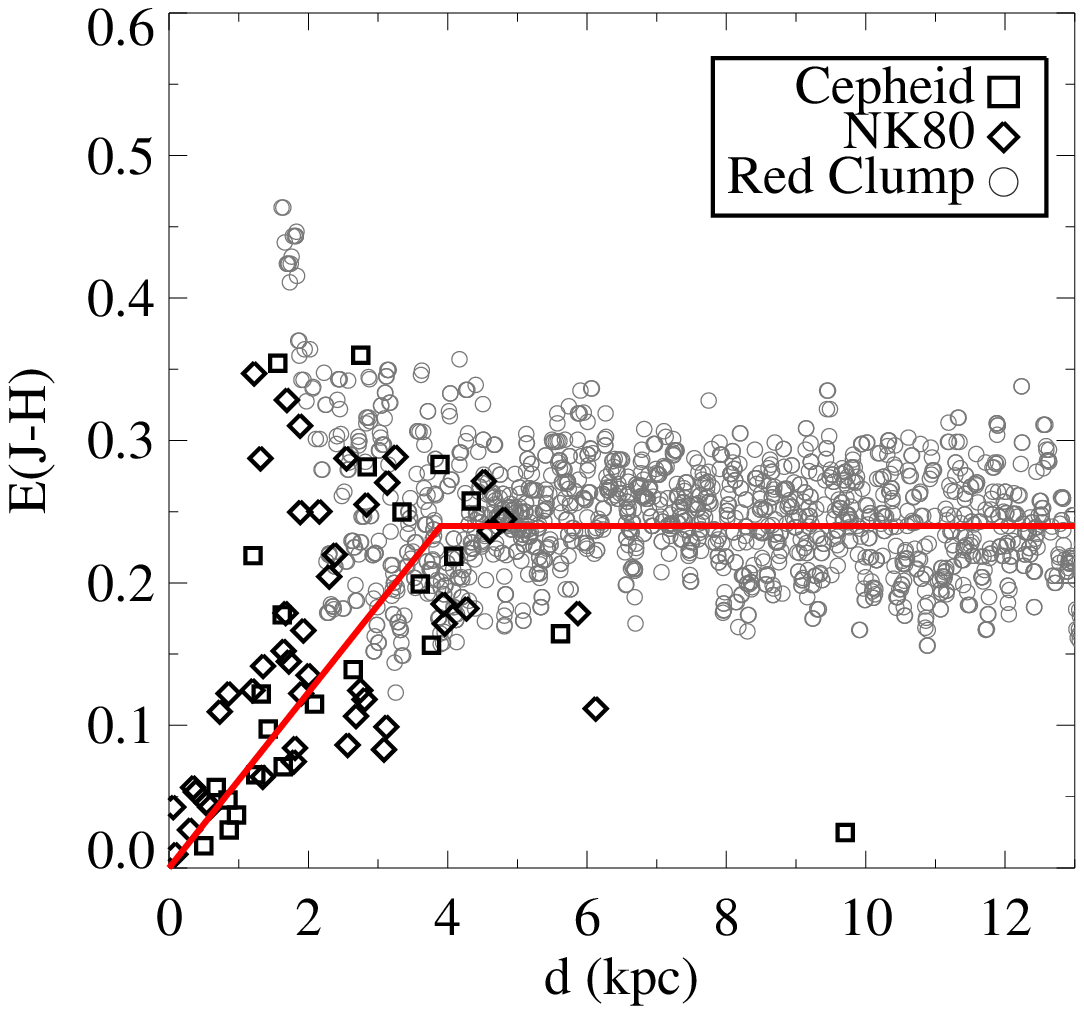} 
\caption{\small{The reddening-distance profile for the sight-line encompassing Andrews-Lindsay 1.  Classical Cepheids \citep{be00,be08} and results from \citet[][NK80]{nk80} were used to trace the local region, while red clump giants were employed to extend the distance baseline.  The mean color-excess for remote sources ($E_{J-H}=0.24\pm0.03 \rightarrow E_{B-V}\sim0.72$) agrees with estimates from optical analyses of Andrews-Lindsay 1, to within the uncertainties (Table~\ref{table1}).}}
\label{fig-ce}
\end{center}
\end{figure}

The first comprehensive photometric analysis of the cluster was likely undertaken by \citet{jp94}, who obtained multiband $BVI_c$ observations from the CTIO 0.9-m telescope.  The cluster parameters established were $d=7.57$ kpc and $E(B-V)=0.72$, as  deduced in part from an inspection of the red clump demographic.  \citet{ca95} obtained deeper $BV$ photometry from La Silla that fostered an improved assessment, and yielded $d\sim11.8$ kpc, $E(B-V)\sim0.75$, and $\tau\sim0.7$ Gyr.  \citet{ca95} further remarked that cluster stars may exhibit subsolar abundances, a suggestion confirmed by subsequent abundance measurements determined using the Blanco 4-m telescope (${\rm [Fe/H]}=-0.51\pm0.30$, \citealt{fr04}).

\citet{cm04} expanded the \citet{ca95} analysis by acquiring SAAO $I_c$ observations, along with new $BV$ photometry.  Those SAAO data are independent of the photometric zero-point established by \citet{jp94}.  The \citet{jp94} observations were used as local standards by \citet{ca95}.  An independent evaluation is important since \citet{cm04} discovered a sizable offset relative to the \citet{jp94} photometry: $\Delta (B-V) = -0.053 \pm0.080$ and $\Delta V=-0.098\pm0.049$.  \citet{cm04} derived the cluster reddening via a $(B-V)$ vs.~$(V-I_c)$ color-color analysis \citep{mc96}, and concluded that Andrews-Lindsay 1 is characterized by $E(B-V)=0.7\pm0.2$, $d=12\pm1$ kpc, and $\tau=800$ Myr.  Generally, prior estimates cited in the literature did not feature uncertainties, and were typically deemed approximate.  \citet{fr04} obtained a radial velocity estimate from two cluster stars ($RV=40\pm10$ km/s), and \citet{fr04b} cite general parameters for Andrews-Lindsay 1 of $d=9.35$ kpc and $\tau=0.67$ Gyr (their Table~1).  Conversely, \citet{ca05} acquired new CTIO $BVI_c$ photometry and deduced $d=16.95$ kpc, $E(B-V)=0.34\pm0.05$, and $\tau=800\pm200$ Myr (their Table~4).  Note that an underestimated $E(B-V)$ can result in an overestimated distance.

\citet{fr08} analyzed SAAO observations and discovered that the PN and cluster velocities are comparable, namely $51.6\pm15.0$ km/s and $50\pm10$ km/s accordingly.  Moreover, \citet{fr08} used the $H\alpha$ SB-r relation to determine a distance for PHR 1315-6555 ($d=9.7\pm3.1$ kpc, $\sigma/d\sim30$\%) which is consistent with estimates for the cluster, e.g., by \citet[][$d=7.54$ kpc]{jp94} and \citet[][$d=12\pm1$ kpc]{cm04}.  \citet{pa11} bolstered the \citet{fr08} findings by acquiring AAOmega data, which imply velocities for the PN of $58.0\pm2.5$ km/s (six Paschen lines) and $59\pm2$ km/s ($H\beta$ and [OIII]).   A coauthor (Frinchaboy) on the \citet{pa11} study determined that 3 cluster stars exhibit a mean velocity of $57 \pm 5$ km/s, which presumably supersedes the earlier \citet{fr04} estimate (Table~\ref{table1}).  \citet{pa11} note that the preliminary SAAO data discussed by \citet{fr08} indicate a PN reddened by $E(B-V)=0.71$, while new high-quality ANU data increase the result to $E(B-V)=0.83\pm0.08$.  The latter was paired with other parameters to establish a revised $H\alpha$ SB-r PN distance of $d=10.4\pm3.4$ kpc, which is marginally larger than the initial \citet{fr08} estimate.  \citet{pa11} conclude that consistent distances, velocities, and extinction estimates for the PN and (mean) cluster indicate the two are associated.  However, an independent assessment of the cluster is desirable given the scatter tied to the estimated distance and color-excess.  To that end \citet{pa11} examined $JHK_s$ photometry from 2MASS, but concluded that the bulk of the cluster lies beyond the survey limits (Fig.~\ref{fig-nircmd}).  Furthermore, it was remarked that faint 2MASS photometry could be contaminated owing to the cluster's compact nature \citep[e.g.,][]{ma12a,ma12b}.  In the following section the fundamental parameters of the cluster are resolved using deeper high-resolution photometry.

\subsection{{\rm \footnotesize EXTINCTION AND DISTANCE}}
The first step to secure the cluster distance was a determination of the reddening along $\ell\sim305 \degr$, particularly given the spread among existing estimates.  The new near-infrared data acquired for Andrews-Lindsay 1 contain numerous red clump giants, which enable the run of reddening to be established across a sizable distance baseline.  Those observations were obtained via the Osiris instrument on the SOAR 4-m telescope (Cerro Pach\'on, Chile).  Standard IRAF and DAOPHOT \citep{st87} routines were employed to extract the photometry.  The instrumental photometry was subsequently standardized using comparatively uncrowded 2MASS stars in the broader field.   

Red field stars were isolated in the near-infrared color-magnitude diagram ($J$ vs.~$J-H$), and assigned an absolute magnitude and intrinsic color tied to a red clump giant \citep{ma11}.  Mean intrinsic parameters for a red clump star were inferred by pairing revised Hipparcos parallaxes \citep{vl07} with the \citet{sk13} catalog of spectral classifications \citep{ma11}.  A distance was computed using the standard extinction law, as deduced by Majaess et al.~(2014, in preparation) from a mapping of the broader region encompassing $\ell\sim305 \degr$.  That analysis relies on WISE mid-infrared data \citep{wr10} to determine extinction law variations via the color-extrapolation method.  For comparison, certain regions in Carina and the Galactic bulge adhere to an anomalous extinction law \citep{pi12,ca13}.

The binned results of $\sim1800$ potential red clump stars surrounding Andrews-Lindsay 1 are shown as Fig.~\ref{fig-ce}.  The findings were supplemented with data from \citet{nk80}, who delineated the run of reddening with distance for numerous Galactic sight-lines using MK spectra and $H\beta$ observations.  The color-excess and distance inferred from classical Cepheids occupying the broader region were likewise added to Fig.~\ref{fig-ce} \citep[][and unpublished observations]{be00,be08}.  The Cepheid distances were established via the latest Galactic Wesenheit $VI_c$ function \citep{ma13}, which relies partly on the HST parallaxes of \citet{be07} and the cluster Cepheids tabulated by \citet{tu10} \citep[see also][]{ng13}. The $VI_c$ function was utilized owing to its diminished sensitivity to abundance variations \citep{ma11b}.  The color-excess was computed with the period-color relation from \citet{ma09} \citep[see also][]{tu01}.

\begin{figure}[!t]
\begin{center}
\includegraphics[width=8.4cm]{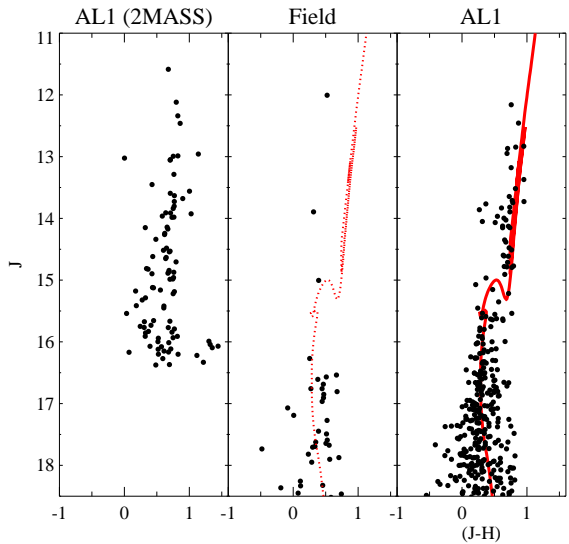} 
\caption{\small{Near-infrared color-magnitude diagrams for Andrews-Lindsay 1.  Fundamental parameters established for the cluster from the new data (right panel) are: $d=9.9\pm0.6$ kpc and $\log{\tau}=8.90\pm0.15$.}}
\label{fig-nircmd}
\end{center}
\end{figure}

The results of Fig.~\ref{fig-ce} imply that the color-excess increases in tandem with distance to $d\sim 4$ kpc (see also Fig.~\ref{fig-vd}), whereupon the trend becomes nearly constant.  An inspection of the local spiral map \citep[][their Fig.~2]{ma11} indicates that the Sagittarius-Carina arm is a source of extinction.  The mean reddening is $E(J-H)=0.24\pm0.03$, and shall help constrain the distance and age of Andrews-Lindsay 1.  The corresponding optical reddening is $E(B-V)\sim0.72$, which agrees with the cluster and PN color-excess advocated by \citet{cm04} and \citet{pa11}, respectively (Table~\ref{table1}).  The \citet{sf11} recalibration of the SFD dust maps yields $E(J-H)\sim0.27$ for the sight-line.\footnote{NED extinction calculator: \url{http://ned.ipac.caltech.edu/}}

\begin{figure}[!t]
\begin{center}
\includegraphics[width=7.5cm]{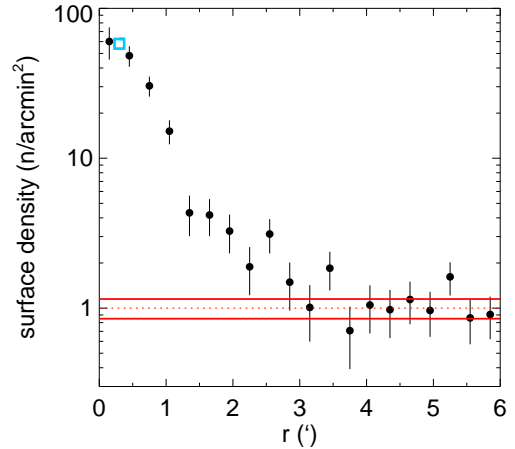} 
\caption{\small{A radial profile constructed from new $BV$ photometry for the region encompassing Andrews-Lindsay 1.  The PN PHR 1315-6555 (blue square) lies well within the bounds of the cluster, which exhibits a corona that extends beyond $r\sim2 \arcmin$.}}
\label{fig-sc}
\end{center}
\end{figure}

\begin{figure}[!t]
\begin{center}
\includegraphics[width=7.5cm]{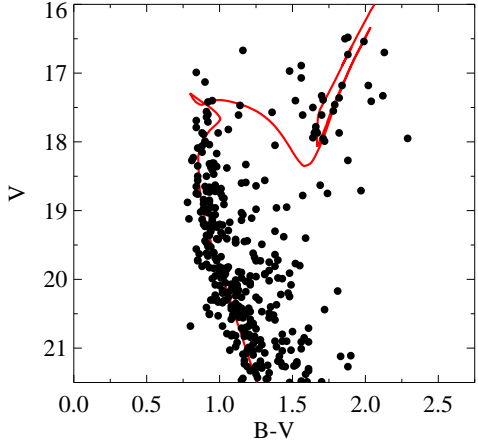} 
\caption{\small{A $BV$ color-magnitude diagram of Andrews-Lindsay 1 implies a cluster distance of $d=10.1\pm0.5$ kpc, which supports the near-infrared solution (Fig.~\ref{fig-nircmd}).  The observations are standardized to new photometry acquired for the field from the CTIO 1-m telescope.}}
\label{fig-bv}
\end{center}
\end{figure}

The cluster distance may be ascertained by shifting an isochrone along the ordinate of the ($JH$) color-magnitude diagram (Fig.~\ref{fig-nircmd}), given the color-excess has been constrained above.  A Padova isochrone \citep{gi02,bo04} was overlaid upon the data, with the aim of matching evolved blue and red stars, in concert with the upper-most main-sequence.  Stars near the cluster core were examined to mitigate field contamination \citep[Fig.~\ref{fig-vd}, and see also Fig.~1 in][]{ca95}.  The fit implies a distance of $d=9.9\pm0.6$ kpc, and an age of $\log{\tau}=8.90\pm0.15$.  A comparison field presented in Fig.~\ref{fig-nircmd} does not follow the isochrone. A field star decontamination algorithm \citep[][and references therein]{bb07} applied corroborates the results, whereby the morphology displayed in the color-magnitude diagram by a field sample was removed from a diagram featuring field and cluster stars.  Metallicity effects in the near-infrared are not acute \citep{al96,ma11c}, and thus a solar isochrone was adopted.  The Padova isochrone was adjusted to match the \citet{ma11c} distance scale, which is not linked to the Hipparcos results for the Pleiades, but rather to 7 other comparatively nearby benchmark clusters that exhibit consistent $JHK_s$ and revised Hipparcos distances \citep{vl09,ma11c}.  The Hyades and Praesepe are two of those clusters, and share a similar age with Andrews-Lindsay 1.  The intrinsic $JHK_s$ relations are tied to revised Hipparcos parallaxes for stars within $d_{\sun}=25$ pc \citep{vl07}, however, the principle concern promulgating the literature pertains to Hipparcos parallaxes for certain further targets \citep[e.g., the Pleiades, Blanco 1, AQ Pup,][see also \citealt{vl09,vl13} and discussion therein]{ma11c,tu12}.  The new cluster parameters constrain the PN's progenitor mass ($2.3 M_{\sun}$), and were paired with observations of the PN \citep[][their Table~5]{pa11} to deduce its radius ($0.3$ pc).  That progenitor mass assumes single star evolution, which may be presumptuous for PNe \citep{ja13,dm13}.  An expansion velocity and observations related to the central star (PN) are needed to glean further pertinent information.  

New $UBVI_c$ observations were acquired via the Y4CAM on the CTIO 1-m telescope to delineate the cluster's radial profile, and check the infrared distance.   Photometry was extracted using DAOPHOT, and subsequently standardized to 46 stars in the \citet{la92} SA~98 and PG~1047 fields.  The new Y4CAM photometry span a sizable field of view, which fosters a reliable determination of the cluster's apparent size.  The region occupied by cluster stars in the color-magnitude diagram was isolated to mitigate field contamination, and to improve the robustness of the aforementioned determination.  The resulting binned analysis implies that Andrews-Lindsay 1 features a corona that extends beyond $r\sim2 \arcmin$ (Fig.~\ref{fig-sc}). PHR 1315-6555 is $r\sim23\arcsec$ from the center of Andrews-Lindsay 1, which in tandem with the suite of evidence presented thus far, is further indication that the PN is associated with the cluster.

\begin{figure}[!t]
\begin{center}
\includegraphics[width=7.5cm]{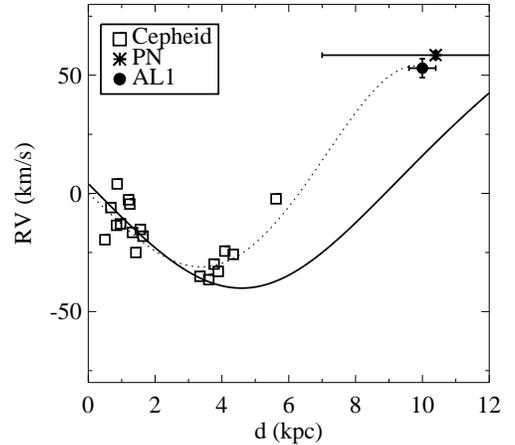} 
\caption{\small{The velocity-distance profile for the region encompassing Andrews-Lindsay 1.  The trend predicted from a simple model for Galactic rotation is denoted by the solid line, whereas  Cepheids were used to delineate the empirical profile \citep[e.g.,][]{be00}.  The predicted relation is generally consistent with the observations, whereby nearer objects exhibit negative velocities and distant targets feature positive velocities.  The spatial and kinematic agreement between the remote PN PHR 1315-6555 and the cluster Andrews-Lindsay 1 imply the former is probably a cluster member.  Deviations from the predicted velocity-distance relation are expected, and a marginal offset appears relative to the empirical trend approximated by the dotted line (fit to the Cepheids).  However, it is the steep gradient of the relation that is (most) important.}}
\label{fig-vd}
\end{center}
\end{figure}

The new Y4CAM observations were employed to re-calibrate the deeper $BV$ photometry of \citet{ca95}, which were originally tied to the \citet{jp94} zero-point.  The reddening cited earlier was used to constrain an optical-based distance ($d=10.1\pm0.5$ kpc, Fig.~\ref{fig-bv}). The result agrees with the near-infrared solution, and a weighted mean yields a distance of $d=10.0\pm0.4$ kpc.  That estimate implies that PHR 1315-6555 features among the most precise distances for PNe ($4$\%, see also \citealt{be09}). 

The aforementioned parameters were determined by matching the observed and intrinsic data via the traditional visual approach \citep[e.g.,][]{cm04,bb10}, whereby the uncertainty represents the limit where a mismatch is clearly perceived. The traditional approach is subjective, and thus a complementary analysis was undertaken using the Bonatto (2014, in preparation) computer algorithm for fitting isochrones (multi-dimensional minimization).  That code yielded a consistent solution of $d=10.20\pm0.52$ kpc, $E(B-V)=0.68\pm0.04$, and $\tau=810\pm100$ Myr (for Z=0.019).  \citet{pn06} remarked that errors tied to isochrone fitting via computer algorithms are comparable to those associated with the canonical approach, which appears supported by the present analysis.  

A simplistic Galactic rotation model was employed to predict the velocity-distance profile for $\ell =305 \degr$, and the results are shown as Fig.~\ref{fig-vd}.  Andrews-Lindsay 1, PHR 1315-6555, and several Cepheids adhere to the general underlying trend, at least to first-order.  Comparatively nearby objects exhibit negative velocities, whereas distant targets display large positive velocities (i.e., the PN and cluster pair).  The magnitude of the velocity-distance gradient is especially pertinent, since it implies that two objects along the sight-line which feature similar velocities may be nearly equidistant.   In other words, the steep slope indicates that two remote objects sharing comparable velocities are at a common distance.  A pronounced gradient likewise mitigates uncertainties arising from imprecise velocity determinations (Table~\ref{table1}).  Admittedly, the PN and cluster, in concert with the more remote Cepheids, appear systematically offset from the predicted velocity-distance correlation.  However, various uncertainties promulgate into the aforementioned determination \citep[e.g., distance to the Galactic center,][]{ma10}, and deviations from simple Galactic rotation are well-known. Andrews-Lindsay 1 is also an older open cluster.  Anthropic arguments aside, Fig.~\ref{fig-vd} provides evidence in favor of an association between PHR 1315-6555 and Andrews-Lindsay 1.  In this instance the probability of detecting two unrelated remote objects that exhibit consistent parameters ($r$, $d$, $RV$) is small.

\section{{\rm \footnotesize CONCLUSION}}
New multiband ($BVJHW_{1-4}$) observations for Andrews-Lindsay 1 and the surrounding field were analyzed, with the objective of highlighting the correct set of cluster parameters (Table~\ref{table1}) and bolstering the link to PHR 1315-6555.  Red clump stars identified in the near-infrared photometry were exploited to constrain the run of reddening with distance for the $\ell \sim 305 \degr$ sight-line (Fig.~\ref{fig-ce}).  Those data were supplemented with information gleaned from Cepheid variables.  The observations imply that the bulk of the reddening is foreground to Andrews-Lindsay 1, and is characterized by $E(J-H)=0.24\pm0.03$.  The reddening along $\ell \sim 305 \degr$ increases in concert with distance until $\sim 4$ kpc, and thereafter remains nearly constant (Figs.~\ref{fig-ce}, \ref{fig-vd}).  The corresponding optical color-excess matches that cited for the PN \citep{fr08,pa11}, to within the uncertainties (Table~\ref{table1}).  

The distance for Andrews-Lindsay 1 was inferred from a solar isochrone fit to cluster stars in color-magnitude ($BVJH$) diagrams (Figs.~\ref{fig-nircmd} and \ref{fig-bv}, $d_{JH}=9.9\pm0.6$ kpc, $d_{BV}=10.1\pm0.5$ kpc, and $\log{\tau}=8.90\pm0.15$).  The mean distance derived lies between the \citet{jp94} and \citet{cm04} estimates.  A velocity-distance correlation predicted for $\ell \sim 305 \degr$ from Galactic rotation generally agrees with the empirical trend delineated by Cepheids, PHR 1315-6555, and Andrews-Lindsay 1.  Specifically, the former exhibit negative velocities conducive to nearer targets, whereas the latter two phenomena feature large positive velocities tied to distant objects (Fig.~\ref{fig-vd}).  The steep correlation indicates that remote objects that are kinematically and spatially coincident along $\ell\sim305\degr$ are likely associated.  The suite of evidence favors an association between the PN PHR 1315-6555 and open cluster Andrews-Lindsay 1 \citep{pa06,ma07,fr08,pa11}.

\subsection*{{\rm \footnotesize ACKNOWLEDGEMENTS}}
\scriptsize{DM is grateful to the following individuals and consortia whose efforts, advice, or encouragement enabled the research: G. Jacoby, D. Frew, A. Parker, P. Frinchaboy, 2MASS, P. Stetson (DAOPHOT), G. Cibis, F. van Leeuwen, F. Benedict, L. Kiss, W. Gieren, D. Balam, B. Skiff, L. Gallo, R. Thacker, Webda (E. Paunzen), W. Dias, CDS (F. Ochsenbein, T. Boch, P. Fernique), arXiv, and NASA ADS.

This publication makes use of data products from the Wide-field Infrared Survey Explorer, which is a joint project of the University of
California, Los Angeles, and the Jet Propulsion Laboratory/California Institute of Technology, funded by NASA; observations
obtained at the Southern Astrophysical Research (SOAR) telescope (program ID: CN2013A-157),  which is a joint project of the Minist\'{e}rio da Ci\^{e}ncia, Tecnologia, e Inova\c{c}\~{a}o (MCTI) da Rep\'{u}blica Federativa do Brasil, the U.S. National Optical Astronomy Observatory (NOAO), the University of North Carolina at Chapel Hill (UNC), and Michigan State University (MSU).}

\end{document}